\documentclass[aps,prd,twocolumn,nofootinbib,showpacs]{revtex4}

\usepackage{graphicx}
\usepackage{color}
\usepackage[colorlinks=true, linkcolor=blue, citecolor=blue, urlcolor=blue]{hyperref}

\begin{document}

\title{The Importance of Proper Renormalization Scale-Setting for Testing QCD at Colliders}

\author{Xing-Gang Wu$^1$}
\author{Sheng-Quan Wang$^{1,2}$}
\author{Stanley J. Brodsky$^3$}

\address{$^1$ Department of Physics, Chongqing University, Chongqing 401331, P.R. China \\
$^2$ School of Science, Guizhou Minzu University, Guiyang 550025, P.R. China \\
$^3$ SLAC National Accelerator Laboratory, Stanford University, Stanford, California 94039, USA}

\date{\today}

\begin{abstract}

A primary problem for perturbative QCD analyses is how to set the renormalization scale of the QCD running coupling in order to achieve maximally precise fixed-order predictions for physical observables. The Principle of Maximum Conformality (PMC) eliminates the ambiguities associated with the conventional renormalization scale-setting procedure, giving predictions which are independent of the choice of renormalization scheme. The scales of the QCD couplings and the effective number of quark flavors are set order by order in the pQCD series. The PMC has a solid theoretical foundation, satisfying the standard renormalization group invariance and all of the self-consistency conditions derived from the renormalization group. The PMC scales at each order are obtained by shifting the arguments of $\alpha_s$ to eliminate all non-conformal $\{\beta_i\}$-terms in the pQCD series. The $\{\beta_i\}$ terms are determined from renormalization group equations without ambiguity. One then obtains the correct behavior of the running coupling at each order and at each phase-space point. The PMC reduces in the $N_C \to 0$ Abelian limit to the Gell-Mann-Low method. In this brief report, we summarize the results of our recent PMC applications for a number of collider processes, emphasizing their generality and applicability. A discussion of hadronic $Z$ decays shows that by applying the PMC, one can achieve accurate scheme-independent predictions for the total and separate decay widths at each order without scale ambiguities.  We also show that if one applies the PMC to determine the top-quark pair forward-backward asymmetry at the next-to-next-to-leading order level, one obtains a comprehensive, self-consistent pQCD explanation for the Tevatron measurements of the asymmetry, accounting for the ``increasing-decreasing" behavior observed by D0 collaboration as the $t \bar t$ invariant mass is increased. At lower energies, one can use the angular distribution of heavy quarks to obtain a direct determination of the heavy quark potential. A discussion of the angular distribution of massive quarks and leptons is also presented, including the fermionic part of the two-loop corrections to the electromagnetic form factors. These results demonstrate that the application of the PMC systematically eliminates a major theoretical uncertainty for pQCD predictions, thus increasing the sensitivity of the colliders to possible new physics beyond the Standard Model.



\end{abstract}

\maketitle

\section{Introduction}

A primary problem for perturbative QCD analyses of hadronic processes, such as those studied at a Tau-Charm Factory, is how to systematically set the renormalization scales of the QCD running coupling in order to achieve precise fixed-order predictions for physical observables. If one uses the conventional scale-setting method, one simply guesses a single renormalization scale $(\mu_r)$ for the argument of the QCD running coupling and varies it over an arbitrary range. This method for setting the scale has inherent difficulties. For example, the resulting pQCD predictions depend on the choice of renormalization scheme, in contradiction to the principle of ``renormalization scheme invariance'' -- predictions for physical observables cannot depend on a theoretical convention~\cite{Wu:2013ei}. Moreover, the error estimate obtained by varying $\mu_r$ over an arbitrary range is unreliable, since the resulting variation of the prediction is only sensitive to perturbative contributions involving the pQCD $\beta$-function. The convergence of the series is problematic due to the presence of divergent renormalon terms. In some processes, a large ``$K$-factor" arises, e.g., for $J/\psi$ production at a $\tau$-charm factory. However, one cannot decide whether the large $K$-factor is indeed the property of the process or a false result due to the improper choice of scale. Worse, guessing the renormalization scale generally gives predictions for QED observables which are in contradiction to the experimentally verified, precise predictions obtained using the standard Gell-Mann-Low (GM-L) method~\cite{GellMann:1954fq}.

The Principle of Maximum Conformality (PMC)~\cite{Brodsky:2011ta, Brodsky:2012sz, Brodsky:2012rj, Brodsky:2011ig, Mojaza:2012mf, Brodsky:2013vpa} provides a systematic way to eliminate renormalization scheme-and-scale ambiguities. It has a rigorous theoretical foundation, satisfying renormalization group (RG)-invariance~\cite{Wu:2014iba} and all of the other self-consistency conditions derived from the renormalization group~\cite{Brodsky:2012ms}. The PMC scales at each order are obtained by shifting the arguments of the running coupling to eliminate all nonconformal $\{\beta_i\}$-terms. The resulting scales also determine the correct effective numbers of flavors $n_f$ at each order. The pQCD convergence is automatically improved due to the elimination of the divergent renormalon series. There can be special cases where the $\beta=0$ conformal terms at low orders, are large, leading to large $K$-factors. This indicates that an even higher-order calculation is needed for a reliable pQCD prediction.

The PMC reduces in the $N_c\to 0$ Abelian limit to the GM-L method~\cite{GellMann:1954fq}. The PMC provides the underlying principle for the well-known Brodsky-Lepage-Mackenzie (BLM) approach~\cite{Brodsky:1982gc}, extending the BLM procedure unambiguously to all orders consistent with the renormalization group. An important example of a BLM/PMC application at next-to-leading order (NLO) level is the investigation of semihard processes based on the BFKL approach~\cite{Brodsky:1999je, Hentschinski:2012kr, Zheng:2013uja, Caporale:2015uva}. At the NLO level, the previous BLM predictions are equivalent to the PMC results, since in both cases one only needs to deal with the $\beta_0$-terms which can be unambiguously fixed~\footnote{One needs to ensure that in these BLM predictions, only the $\beta_0$-terms that pertain to $\alpha_s$-running are eliminated.}.

The PMC has now been successfully applied to a number higher-order processes, providing precisions tests at pQCD at a range of experimental facilities, including electron-positron annihilation to hadrons~\cite{Mojaza:2012mf, Brodsky:2013vpa, Wu:2014iba}; Higgs decays to $\gamma\gamma$~\cite{Wang:2013akk}, $gg$ and $b\bar{b}$~\cite{Wang:2013bla, Zeng:2015gha}; hadronic $Z$ decays~\cite{Wang:2014aqa}; $\Upsilon(1S)$ leptonic decay~\cite{Shen:2015cta}; top-pair production at the LHC and Tevatron~\cite{Brodsky:2012sz, Brodsky:2012rj, Brodsky:2012ik, Wang:2014sua, Wang:2015lna}, etc. When it is applied to $B$ physics, the PMC provides a possible solution to the $B\to\pi\pi$ puzzle~\cite{Qiao:2014lwa}.

Measurements of the $Z$-boson decay rate into hadrons can provide an important method for determining a high precision value for the strong coupling constant $\alpha_s$ at a specific renormalization scale. This is the central goal of GigaZ~\cite{Baer:2013cma} and other super-$Z$ factories~\cite{superZ} -- high luminosity $e^+ e^-$ colliders operating at the $Z$-resonance. The contributions from the nonperturbative and power-law terms are suppressed, and the smallness of $\alpha_s$ leads to a rapid decrease of the higher-order corrections in the pQCD series. In fact, by applying PMC scale-setting to the available pQCD predictions up to four-loop level~\cite{Baikov:2012er, Baikov:2008jh, Baikov:2010je, Baikov:2012zn}, one obtains optimal fixed-order predictions for the $Z$-boson hadronic decay rate, thus enabling a very high precision test of the Standard Model~\cite{Wang:2014aqa}.

The authors of Ref.\cite{Czakon:2014xsa} have noted that an alternative scale-setting procedure, called the ``large $\beta_0$-approximation"~\cite{Neubert:1994vb, Beneke:1994qe},
leads to incorrect next-to-next-to-leading order (NNLO) $n^2_f$-term to the top-pair production at hadron colliders. It should be emphasized that this analytic error is a defect of ``large $\beta_0$-approximation"; it does not occur if one uses PMC scale-setting.

It is possible for conventional scale-setting to accidentally predict the correct value of a global observable such as the total cross-section at sufficiently high order; however, since one assumes the same renormalization scale at each order in $\alpha_s$, it will often give incorrect predictions for differential observables. In fact, as in QED, the scale and effective number of flavors is distinct at each order of pQCD, reflecting the different virtualities of the relevant subprocesses as a function of phase space. This provides the underlying reason why a single `guessed' scale cannot explain the ``increasing-decreasing" behavior of $A_{\rm FB}$ as the $t \bar t$-pair mass is varied. We shall show that the PMC provides a self-consistent explanation for all of the $t \bar t$-pair measurements at the Tevatron.

The PMC can also be applied to problems with multiple physical scales. For example, the subprocess $q \bar q \to Q\bar Q$ near the quark threshold involves, not only the subprocess scale $\hat{s} \sim 4 M^2_Q$, but also the scale $v^2 \hat{s}$ which appears in the Sudakov final-state corrections~\cite{Brodsky:1995ds}. Here $v$ is the $Q \bar Q$ relative velocity which becomes very small near threshold, and $Q$ labels the heavy-quark flavor. We need to introduce two PMC scales for this process, one for the hard-part and one for the Coulomb-terms at the presently known order of the pQCD corrections. This is an important PMC application for processes at super tau-charm factories.

The following sections of this contribution to {\it Frontiers of Physics in China} are organized as follows: In Sec.II, we present general arguments for the scale-setting problem and introduce the ${\cal R}_\delta$-scheme, which provides a systematic and convenient way to identify the nonconformal $\beta$ terms in a pQCD series needed to compute the PMC scales. We review several applications which will illustrate important features of the PMC: hadronic $Z$ decay rates; the cross section and the forward-backward asymmetry of top-quark pair production at the Tevatron; and the angular distributions of massive quarks and leptons close to threshold -- which is directly relevant to physics studies at a super tau-charm factory. This PMC application also illustrates how to deal with multiple-scale problems. Sec.III is reserved for the summary.

\section{General Arguments for Proper Scale-Setting and the PMC}

The scale dependence of the running coupling is controlled by the renormalization group equation (RGE), which can be used recursively to establish the perturbative pattern of $\beta$ terms at each order. More explicitly, the scale-displacement relation for the running coupling at two different scales $\mu_1$ and $\mu_2$ defines the following $\beta$-pattern at each order,
\begin{eqnarray}
\alpha(\mu_2) &=& \alpha(\mu_1)- \beta_{0} \ln\left(\frac{\mu_2^{2}} {\mu_1^2}\right) \alpha^{2}(\mu_1) \nonumber\\
&& +\left[\beta^2_{0} \ln^2 \left(\frac{\mu_2^{2}}{\mu_1^2}\right) -\beta_{1} \ln \left(\frac{\mu_2^{2}} {\mu_1^2}\right)\right] \alpha^{3}(\mu_1)  + \ldots ,  \label{scaledis}
\end{eqnarray}
where $\alpha=\alpha_s/4\pi$. The PMC utilizes this perturbative $\beta$-pattern to systematically set the scales of the running coupling at each order in a pQCD expansion; the coefficients of the resulting series thus match the coefficients of the corresponding conformal theory with $\beta=0$. This is the same principle used in QED where all $\beta$-terms resulting from the vacuum polarization corrections to the photon propagator are absorbed into the scale of the running coupling. As in QED, the scales are physical in the sense that they reflect the virtuality of the gluon propagators at a given order, as well as setting the effective number $n_f$ of active flavors. The resulting resummed pQCD expression thus determines the relevant ``physical" scales for any physical observable, thereby providing a solution to the renormalization scale-setting problem.

We have suggested two all-orders PMC approaches which are equivalent to each other at the level of conformality, and are thus equally viable PMC procedures~\cite{Bi:2015wea}. In this report, we introduce the PMC approach using the ${\cal R}_\delta$-scheme, which can be readily automatized. The ${\cal R}_\delta$ method uses the $\beta$-pattern generated by the RGE and its degeneracy relations to identify which terms in the pQCD series are associated with the QCD $\beta$-function and which terms remain in the $\beta=0$ conformal limit. The $\beta$-terms are then systematically absorbed by shifting the scale of the running coupling at each order, thus providing the PMC scheme-independent prediction.

The starting point of this approach is to introduce an arbitrary dimensional renormalization scheme: the $R_\delta$-scheme. In the $R_\delta$-scheme, an arbitrary constant $-\delta$ is subtracted in addition to the standard subtraction $\ln 4 \pi - \gamma_E$ for the $\overline{\rm MS}$-scheme. This amounts to redefining the renormalization scale by an exponential factor, $\mu_\delta = \mu_{\overline{\rm MS}} \exp(\delta/2)$. The $\delta$-subtraction thus defines an infinite set of new renormalization schemes. Using the ${\cal R}_\delta$-scheme, one can determine the $\beta$-pattern at each perturbative order~\cite{Mojaza:2012mf, Brodsky:2013vpa}. The QCD prediction $\varrho_n$ of a physical observable $\varrho$ can be expressed as
\begin{eqnarray}
\varrho_{n}(Q) &= & r_{0,0} + r_{1,0} \alpha(\mu_r) + \left[r_{2,0} + \beta_0 r_{2,1} \right] \alpha^2(\mu_r) + \nonumber\\
&& \left[r_{3,0} + \beta_1 r_{2,1} + 2 \beta_0 r_{3,1} + \beta _0^2 r_{3,2} \right] \alpha^3(\mu_r) +\cdots, \label{betapattern}
\end{eqnarray}
where $Q$ stands for the scale at which it is measured, all the coefficients $r_{i,j}$ are functions of the initial choice of scale $\mu_r$ and $Q$. The $r_{i,0}$ are the conformal parts of the coefficients. Here the $\beta$-pattern for the pQCD series at each order is a superposition of all of the $\{\beta_i\}$-terms which govern the evolution of the lower-order $\alpha_s$ contributions at this particular order.

After applying the standard scale-setting procedures, by setting the PMC scales $Q_i$, the final pQCD prediction for $\varrho_n$ reads
\begin{equation}
\varrho_n(Q) = r_{0,0} + \sum_{i=1}^{n} r_{i,0} \alpha^{i}(Q_i).
\end{equation}
The PMC scales $Q_i$ and the conformal coefficients $r_{i,0}$ can be found in Ref.\cite{Brodsky:2013vpa}. The PMC scales $Q_i$ are functions of $\mu_r$ and $Q$.
The resulting values of the $Q_i$ are independent of the choice of the initial $\mu_r$ at infinite order, ensuring the renormalization group invariance. Thus one can adopt any initial value of $\mu_r$ (only needs to be in perturbative region) and obtain the same pQCD prediction. At fixed orders, there can be a small residual scale-uncertainty in $Q_i$ and hence the final pQCD expression due to the truncation of the $\beta$-function; however, these residual uncertainties are found to be highly suppressed -- in fact negligible -- even for lower-order predictions. Thus, the conventional renormalization scale dependence has been eliminated. In practice, one can take the usual choice of scale; e.g., the typical momentum flow of the process, as the initial scale to simplify the pQCD expressions and the PMC treatment.

The PMC eliminates the renormalization scheme-and-scale dependences which characterize the conventional scale setting methods. It does not break other properties of the pQCD series, such as gauge invariance. Moreover, the PMC correctly sets the scales at each order of pQCD; thus in distinction to conventional scale setting, it simultaneously predicts the correct values for both the total cross-section (or total decay width) and the differential observables.

\subsection{The higher-order pQCD corrections to hadronic $Z$ decays}

The decay rate of the $Z$-boson into hadrons can be expressed as,
\begin{displaymath}
\Gamma_Z = {\cal C}\bigg[\sum\limits_{f}v^2_f r^V_{\rm NS}+ \bigg(\sum\limits_{f}v_f\bigg)^2 r^V_S +\sum\limits_{f} a^2_f r^A_{\rm NS}+r^A_S\bigg],
\end{displaymath}
where the factor ${\cal C}=3 \frac{G_{F}M^3_Z} {24\pi\sqrt{2}}$, $v_f\equiv 2I_f-4q_fs^2_W$, $a_f\equiv2I_f$, $q_f$ is the $f$-quark electric charge, $s_W$ is the effective weak mixing angle, and $I_f$ is the third component of weak isospin of the left-handed component of $f$. $r^V_{\rm NS}=r^A_{\rm NS}\equiv r_{\rm NS}$, $r^V_S$ and $r^A_S$ stand for the non-singlet, vector-singlet and axial-singlet part, respectively. These contributions can be further expressed as
\begin{displaymath}
r_{\rm NS} = 1+\sum^n_{i=1}C^{\rm NS}_ia^i_s, r^V_S =\sum^n_{i=3}C^{\rm VS}_{i}a^i_s, ~r^A_S=\sum^n_{i=2}C^{\rm AS}_{i}a^i_s ,
\end{displaymath}
where $a_s=\alpha_s/\pi$. The coefficients of $r_{\rm NS}$, $r^V_S$ and $r^A_S$ with their dependence on the choice of initial scale can be derived from Refs.\cite{Baikov:2012er, Baikov:2008jh, Baikov:2010je, Baikov:2012zn}. Each physical contribution to the $Z$ decay has a different momentum flow, thus, the PMC scales for $r_{\rm NS}$, $r^V_S$ and $r^A_S$ should be set separately~\cite{Wang:2014aqa}. The prediction for a physical observable should not depend on scheme or the initial choice of the scale. We shall take the non-singlet contribution $r_{\rm NS}$ as example to show that the computed PMC scales and thus the final PMC predictions for the hadronic $Z$ decay results are highly independent of the scale.

\begin{widetext}
\begin{center}
\begin{figure}[htb]
\includegraphics[width=0.48\textwidth]{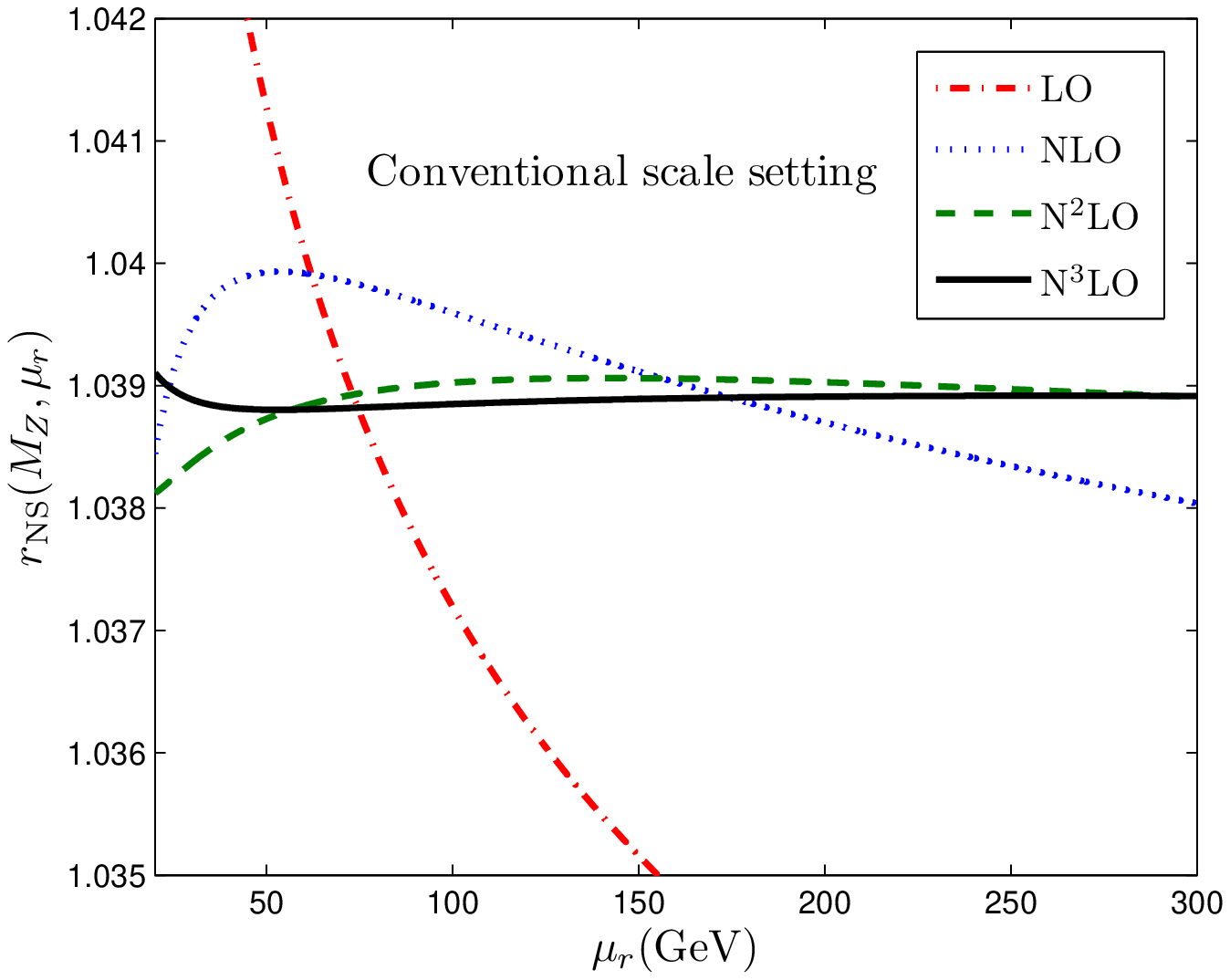}
\includegraphics[width=0.48\textwidth]{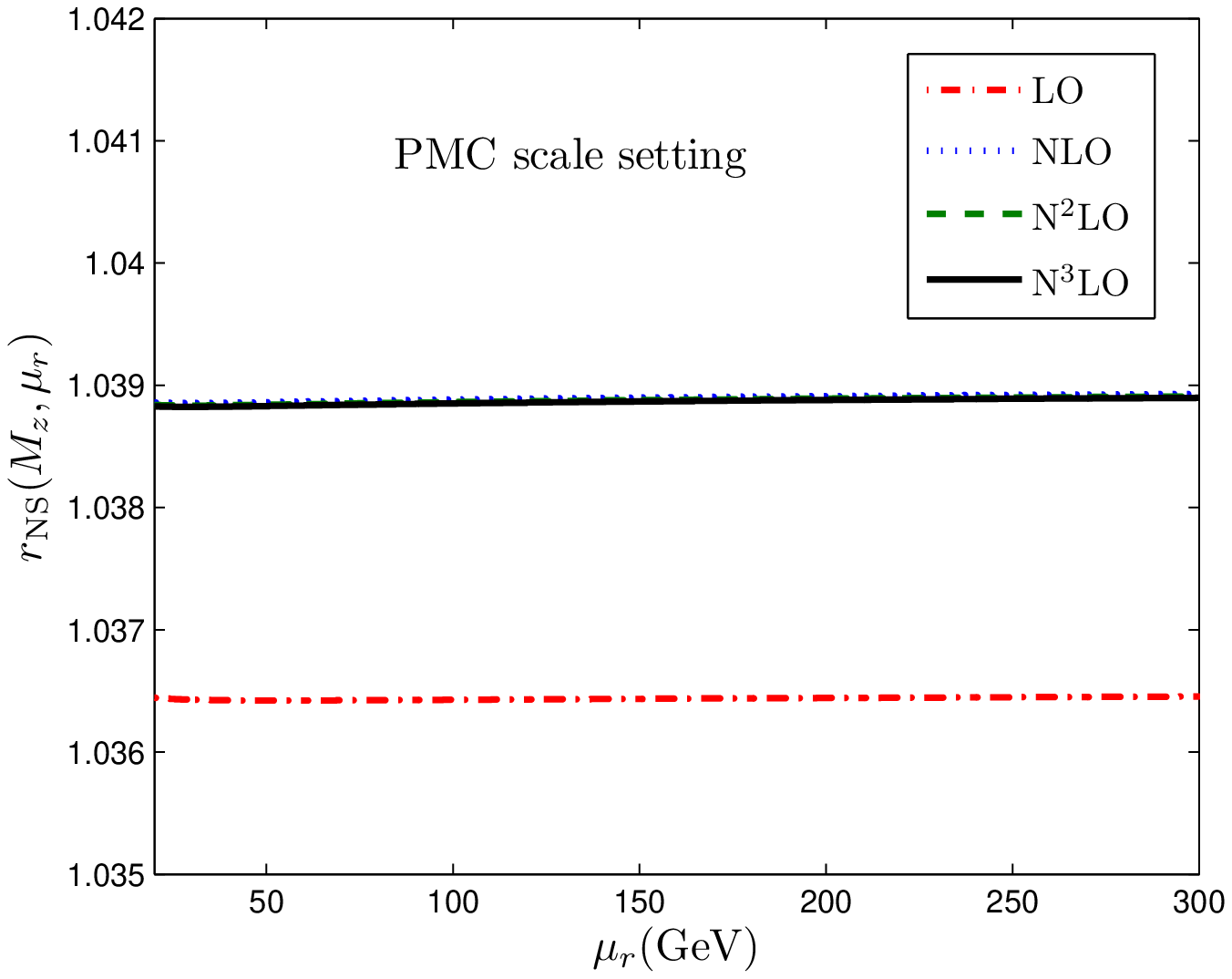}
\caption{The non-singlet contribution $r_{\rm NS}$ up to four-loop QCD corrections versus $\mu_r$ before and after PMC scale setting. After PMC scale setting, the curves for NLO, N$^2$LO, and N$^3$LO are almost coincide with each other.}
\label{Rnscon_pmc}
\end{figure}
\end{center}
\end{widetext}

We present the scale dependence before and after PMC scale setting for $r_{\rm NS}$ in Fig.(\ref{Rnscon_pmc}). As expected, for the case of conventional scale setting, the resulting low-order predictions depend heavily on $\mu_r$, but as expected one observes that as more loops are taken into consideration, a weaker scale dependence is achieved. On the other hand, after applying the PMC, the pQCD predictions at each order are almost independent to $\mu_r$. This is because that the PMC scales at each order are determined unambiguously by absorbing all non-conformal $\beta$-terms into the running coupling. It also indicates that the PMC predictions have the property that any residual dependence on the choice of initial scale is highly suppressed even for low-order predictions. Fig.(\ref{Rnscon_pmc}) shows not only that the renormalization scale ambiguities are eliminated, but also that the value of $r_{\rm NS}$ rapidly approaches its steady point; i.e; the curves at NLO, N$^2$LO, and N$^3$LO almost coincide with each other after applying the PMC.

\begin{table}[htb]
\centering
\begin{tabular}{|c|c|c|c|c|c|}
\hline
~~~ ~~~ & $R^{\rm NS}_{1}$ & $R^{\rm NS}_{2}$ & $R^{\rm NS}_{3}$ & $R^{\rm NS}_{4}$  &~$\sum^{4}_{i={1}} R^{\rm NS}_{i}$~ \\
\hline
Conv. &  0.03769  & 0.00200 & -0.00069 & -0.00016 & 0.03884 \\
\hline
PMC &  0.03636 & 0.00252 & -0.00003 & -0.00001 & 0.03885\\
\hline
\end{tabular}
\caption{Perturbative contributions for the non-singlet $r_{\rm NS}$ under the conventional (Conv.) and the PMC scale settings. Here, $R^{\rm NS}_{i}=C^{\rm NS}_i a^i_s$ with $i=(1,\cdots,4)$ stand for the one-, two-, three-, and four-loop terms, respectively. $\mu_r=M_Z$. } \label{table:each}
\end{table}

We emphasize that after applying the PMC, one obtains better pQCD convergence due to the elimination of the renormalon terms. The pQCD estimations at each perturbative order for $r_{\rm NS}$ are presented in Table \ref{table:each}. The fastest pQCD convergence is thus achieved by applying the PMC. The pQCD correction at ${\cal O}(\alpha^4_s)$ is -0.00016 for the conventional scale setting, which leads to a shift of the central value of $\alpha_s(M_Z)$ from 0.1185 to 0.1190~\cite{Baikov:2008jh, Baikov:2010je}. In contrast, after applying the PMC, this correction becomes a negligible $-0.00001$. Table \ref{table:each} also shows that the predictions for the total sum $\sum^{4}_{i=1} R^{\rm NS}_{i}$ are close in value for both PMC and conventional scale setting, although their perturbative series behave very differently.

\begin{figure}[htb]
\includegraphics[width=0.48\textwidth]{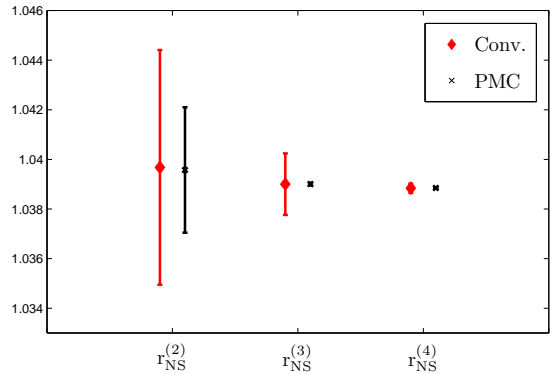}
\caption{The values of $r^{(n)}_{\rm NS}=1+\sum^n_{i=1}C^{\rm NS}_i a^i_s$ and their errors $\pm |C^{\rm NS}_n a^n_s|_{\rm MAX}$. The diamonds and the crosses are for conventional (Conv.) and PMC scale settings, respectively. The central values assume the initial scale choice $\mu_r=M_Z$. } \label{uncert}
\end{figure}

It is helpful to be able to estimate the magnitudes of the ``unknown" higher-order pQCD corrections. One usually estimates those unknown contributions by varying $\mu_r\in[M_Z/2,2M_Z]$. However, this simple procedure only exposes the $\beta$-dependent non-conformal terms -- not the entire perturbative series. We emphasize that after applying the PMC, the scales are optimized and cannot be varied; otherwise, one will explicitly break the renormalization group invariance, leading to an unreliable prediction. Here, for a conservative estimate, we take the uncertainty to be the last known perturbative order~\cite{Wu:2014iba}; i.e. at $n$-th order the perturbative uncertainty is estimated by $\pm |C^{\rm NS}_n a^n_s|_{\rm MAX}$, where the symbol ``MAX" stands for the maximum of $|C^{\rm NS}_n a^n_s|$ by varying $\mu_r$ within the region of $[M_Z/2,2M_Z]$. The error bars for PMC and the conventional scale setting are displayed in Fig.(\ref{uncert}). It shows that the predicted error bars from the ``unknown" higher-order corrections quickly approach their steady points after applying the PMC. These error bars provide a consistent estimate of the ``unknown" QCD corrections under conventional and PMC scale settings; i.e., the exact value for ``unknown" $r^{(n)}_{\rm NS}$ ($n=3$ and $4$) are well within the error bars predicted from $r^{(n-1)}_{\rm NS}$.

\subsection{The yields and the forward-backward asymmetry of the top-pair productions at the Tevatron}

In the following, we summarize our recent results for the yields and the forward-backward asymmetry of the top-pair productions at the Tevatron with hadron-hadron collision energy $\sqrt{S}=1.96$ TeV.

\subsubsection{The $t \bar t$ Production Cross-Section}

The NNLO total cross-sections from all the production channels, i.e. the $(q\bar{q})$-, $(gq)$-, $(g\bar{q})$- and $(gg)$- channels, before and after PMC scale-setting, are~\cite{Wang:2015lna}
\begin{eqnarray}
\sigma_{\rm Total}|_{\rm Conv.} &=& 7.42^{+0.25}_{-0.29} \; {\rm pb}, \label{convcs} \\
\sigma_{\rm Total}|_{\rm PMC}   &\simeq& 7.55 \; {\rm pb},
\end{eqnarray}
where the errors are obtained by varying the initial scale $\mu_r\in [m_t/2, 2m_t]$. The PMC total cross-section is almost unchanged, indicating the residual scale dependence is negligible. Both the PMC and conventional scale-setting procedures agree with the CDF and D0 measurements within errors; the recent combined cross-section given by the CDF and D0 collaborations is $7.60\pm0.41$ pb~\cite{Aaltonen:2013wca}. The dependence of the total cross-section on the choice of renormalization scale is also small using conventional scale-setting if one incorporates NNLO QCD corrections.

\begin{table}[htb]
\centering
\begin{tabular}{|c|c|c|c|c|}
\hline
& \multicolumn{3}{c|}{Conventional} & PMC \\
\hline
~$\mu_r$~  & ~$m_t/2$~  & ~$m_t$~ & ~$2m_t$~ & ~$[m_t/2,2m_t]$~  \\
\hline
~$\sigma^{\rm{LO}}_{q\bar{q}}$~ & ~5.99~ & ~4.90~ & ~4.09~ & ~$\simeq 4.76$~ \\
\hline
~$\sigma^{\rm{NLO}}_{q\bar{q}}$~ & ~0.09~ & ~0.96~ & ~1.41~ & ~$\simeq 1.73$~ \\
\hline
~$\sigma^{\rm{N^2LO}}_{q\bar{q}}$~ & ~0.45~ & ~0.48~ & ~0.63~ & ~$\simeq-0.06$~ \\
\hline
\end{tabular}
\caption{The $(q\bar{q})$-channel cross-sections (in unit: pb) at each perturbative order under the conventional and PMC scale-settings~\cite{Wang:2015lna}, where three typical renormalization scales $\mu_{r}=m_t/2$, $m_t$ and $2m_t$ are adopted. The factorization scale is taken as $\mu_f=m_t$. }
\label{tevat}
\end{table}

Eq.(\ref{convcs}) shows that if one uses conventional scale setting, the scale dependence for the total cross-section at the NNLO level is small; i.e. the scale error is $\left(^{+3\%}_{-4\%}\right)$. However, using a single guessed scale does not predict the cross-sections for individual channels correctly at each order. In fact, by analyzing the pQCD series in detail, we find that the errors for the separate cross-sections at each order from conventional scale-setting are large in all of the contributing channels. As an example, the contributions of the dominant $(q\bar{q})$-channel with and without PMC scale-setting are presented in Table \ref{tevat}. To show the scale dependence of individual cross-sections $\sigma^{i}_{q\bar{q}}$, we define a ratio $\kappa_i$:
\begin{displaymath}
\kappa_i=\frac{\left. \sigma^i_{q\bar{q}}\right|_{\mu_{r}=m_t/2} -\left. \sigma^i_{q\bar{q}}\right|_{\mu_{r}=2m_t}}{\left.\sigma^i_{q\bar{q}} \right|_{\mu_{r}=m_t}},
\end{displaymath}
where $i$=LO, NLO and N$^2$LO, respectively. Using conventional scale-setting, we obtain
\begin{displaymath}
\kappa_{\rm LO}=39\%,~\kappa_{\rm NLO}=-138\%,~\kappa_{\rm N^2LO}=-36\%.
\end{displaymath}
These results show that if one uses conventional scale-setting, then the dependence on the choice of initial scale at each order is very large. For example, the scale dependence of $\sigma^{\rm NLO}_{q\bar{q}}$, which gives the dominant component of the asymmetry $A_{\rm FB}$, reaches up to $-138\%$. On the other hand, by using the PMC, all of the $\kappa_i$-values become less than $0.1\%$, indicating the scale dependence of each loop-term is eliminated simultaneously.

In these calculations, we have set the factorization scale $\mu_f$ to be the renormalization scale $\mu_r$. The determination of the factorization scale is a completely separate issue from the renormalization scale setting, since it is present even for a conformal theory with $\beta=0$. The factorization scale should be chosen to match the nonpertubative bound-state dynamics with perturbative DGLAP evolution, which can be done explicitly by using nonperturbative models such as AdS/QCD and light-front holography, where the light-front wavefunctions of the hadrons are known. See the recent review~\cite{Brodsky:2014yha}. We have found that the factorization scale dependence is decreased after applying the PMC~\cite{Wang:2014sua}. In contrast, the simple conventional scale-setting procedure of setting $\mu_r=m_t$ to eliminate the log-terms $\ln^{k}\mu_r^2/m_t^2$ is again problematic, since it may lead to a large factorization-scale dependence. This again explains the importance of proper renormalization scale setting.

\subsubsection{The $t\bar{t}$ Forward-Backward Asymmetry}

The top-quark pair forward-backward asymmetry in $\bar p p \to t \bar t X$ collisions is also sensitive to the renormalization scale-setting procedure.
This asymmetry is dominated by interference of different amplitudes contributing to the $(q\bar{q})$-channel. Contributions to the asymmetry start at the NLO level. Thus evaluating the correct value for the NLO-terms is very important in order to achieve the correct prediction for the $t \bar t$ asymmetry. In contrast, one cannot trust the value of $\sigma^{\rm NLO}_{q\bar{q}}$ derived using conventional scale-setting due to its large scale errors; i.e., $\kappa_{\rm NLO}\sim -138\%$. Moreover, if one uses conventional scale-setting, the NLO cross-section $\sigma^{\rm NLO}_{q\bar{q}}$ increases and the total cross-section $\sigma^{\rm Total}_{q\bar{q}}$ decreases as $\mu_r$ is increased. Thus, in order to agree with the measured total cross-section, the conventional method requires a smaller scale $\mu_r = m_t/2$; however this leads to the prediction of a small $t \bar t$ asymmetry, well below the data. This example shows why previous NLO SM predictions could not achieve a consistent simultaneous explanation of both the top-pair total cross-section and the $t \bar t$ asymmetry.

\begin{figure}[htb]
\includegraphics[width=0.48\textwidth]{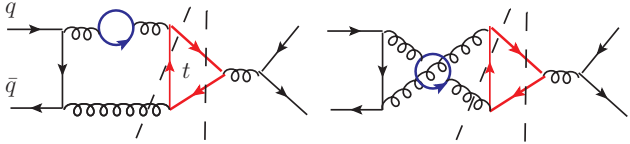}
\caption{Dominant cut diagrams for the $n_f$-terms at the $\alpha^4_s$-order of the $(q\bar{q})$-channel, which are responsible for the smaller effective NLO PMC scale, where the solid circles stand for the light quark loops. }
\label{nloscale}
\end{figure}

It is important to note the NLO PMC scale $\mu^{\rm PMC, NLO}_r$ of the $(q\bar{q})$-channel is much smaller than $m_{t}$. It is dominated by the non-Coulomb $n_f$-terms at the $\alpha^4_s$-order, which are shown in Fig.(\ref{nloscale}). In these diagrams, the momentum flow of the virtual gluons has a large range of virtualities. The NLO PMC scale is numerically small since, in effect, it is a weighted average of the different momentum flows of the gluons. The resulting NLO $(q\bar{q})$ cross-section is in fact about twice as large as the cross-section predicted by conventional scale-setting; the precision of the predicted asymmetry $A_{\rm FB}$ is also greatly improved.

Moreover, as shown by Table \ref{tevat}, after applying the PMC, the predicted ratio of the cross-section at the N$^2$LO level to the NLO cross-section for the $q\bar{q}$-channel, i.e. $|\sigma_{q\bar{q}}^{\rm N^2LO} / \sigma_{q\bar{q}}^{\rm NLO}|$, is reduced from $\sim 50\%$ to less than $\sim 4\%$. This shows that a great improvement of pQCD convergence can be achieved by using the PMC. Such an improvement of the pQCD convergence is essential for achieving accurate pQCD predictions for the $t \bar t$ asymmetry. If we further include the $\mathcal{O}(\alpha^2_s\alpha)$ and the $\mathcal{O}(\alpha^2)$ electroweak contributions, we achieve a precise SM ``NLO-asymmetry" prediction~\cite{Brodsky:2012ik},
\begin{eqnarray} \label{asymmetry}
A^{\rm PMC}_{\rm FB}={\alpha^3_s N_1+\alpha^2_s \alpha \tilde{N}_1 + \alpha^2 \tilde{N}_0 \over \alpha^2_s D_0 + \alpha^3_s D_1},
\end{eqnarray}
where the $D_i$-terms stand for the total cross-sections at each $\alpha_s$-order and the $N_i$-terms stand for the corresponding asymmetric contributions. The term labeled $\tilde{N}_1$ corresponds to the QCD-QED interference contribution at the order ${\cal O}(\alpha^2_s \alpha)$, and $\tilde{N}_0$ stands for the pure electroweak antisymmetric ${\cal O}(\alpha^2)$ contribution arising from $|{\cal M}_{q\bar{q}\to\gamma\to t\bar{t}}+{\cal M}_{q\bar{q}\to Z^0\to t\bar{t}}|^2$. By using Eq.(\ref{asymmetry}), we obtain a precise prediction for $A_{\rm FB}$ without any renormalization scale uncertainty~\cite{Wang:2015lna}, $A^{\rm PMC}_{\rm FB} = 9.2\%$, agreeing with the D0 measurement within errors, $A^{\rm D0}_{\rm FB}=(10.6\pm3.0)\%$~\cite{Abazov:2014cca} and $A^{\rm D0}_{\rm FB}=(11.8\pm2.5\pm1.3)\%$~\cite{Abazov:2015fna}.

We can also use the PMC to predict the top-quark pair asymmetry $A_{\rm FB}(M_{t\bar{t}}>M_{\rm cut})$ as a function of the top-pair invariant mass lower limit $M_{\rm cut}$. In the case of $M_{\rm cut}=450$ GeV, the predicted asymmetry using conventional scale-setting is $A_{\rm FB}(M_{t\bar{t}}>450~\rm GeV)|_{\rm Conv.}=12.9\%$, which becomes even larger after applying the PMC, $A_{\rm FB}(M_{t\bar{t}}>450~\rm GeV)|_{\rm PMC}=29.9\%$. The prediction using conventional scale-setting deviates significantly from the CDF measurements $(47.5\pm11.4)\%$~\cite{Aaltonen:2011kc} and $(29.5\pm5.8\pm3.3)\%$~\cite{Aaltonen:2012it}. In contrast, the PMC prediction agrees with the weighted average of the CDF measurements~\cite{Aaltonen:2011kc, Aaltonen:2012it} within errors. Thus, after applying the PMC, the large discrepancies between the Standard Model estimates and the CDF measurements which were obtained using conventional scale-setting is removed.

\begin{table}[htb]
\centering
\begin{tabular}{|c|c|c|c|c|c|c|}
\hline
~ ~ & \multicolumn{6}{c|}{$M_{\rm cut}$ (GeV)} \\
\cline{2-7}
 $A_{\rm FB}(M_{t\bar{t}}>M_{\rm cut})$ & ~400~ & ~450~ & ~500~ & ~600~ & ~700~ & ~800~ \\
\hline
Conv. & 11\% & 13\% & 15\% & 18\% & 21\% & 23\%  \\
\hline
PMC & 17\% & 30\% & 44\% & 38\% & 31\% & 30\% \\
\hline
\end{tabular}
\caption{Top-quark pair asymmetries $A_{\rm FB}(M_{t\bar{t}}>M_{\rm cut})$ using conventional (Conv.) and PMC scale-setting procedures~\cite{Wang:2015lna}, respectively. The predictions are shown for typical values of $M_{\rm cut}$. The initial scale $\mu_r=m_t$. } \label{tab_mtcut}
\end{table}

\begin{widetext}
\begin{center}
\begin{figure}[htb]
\includegraphics[width=0.48\textwidth]{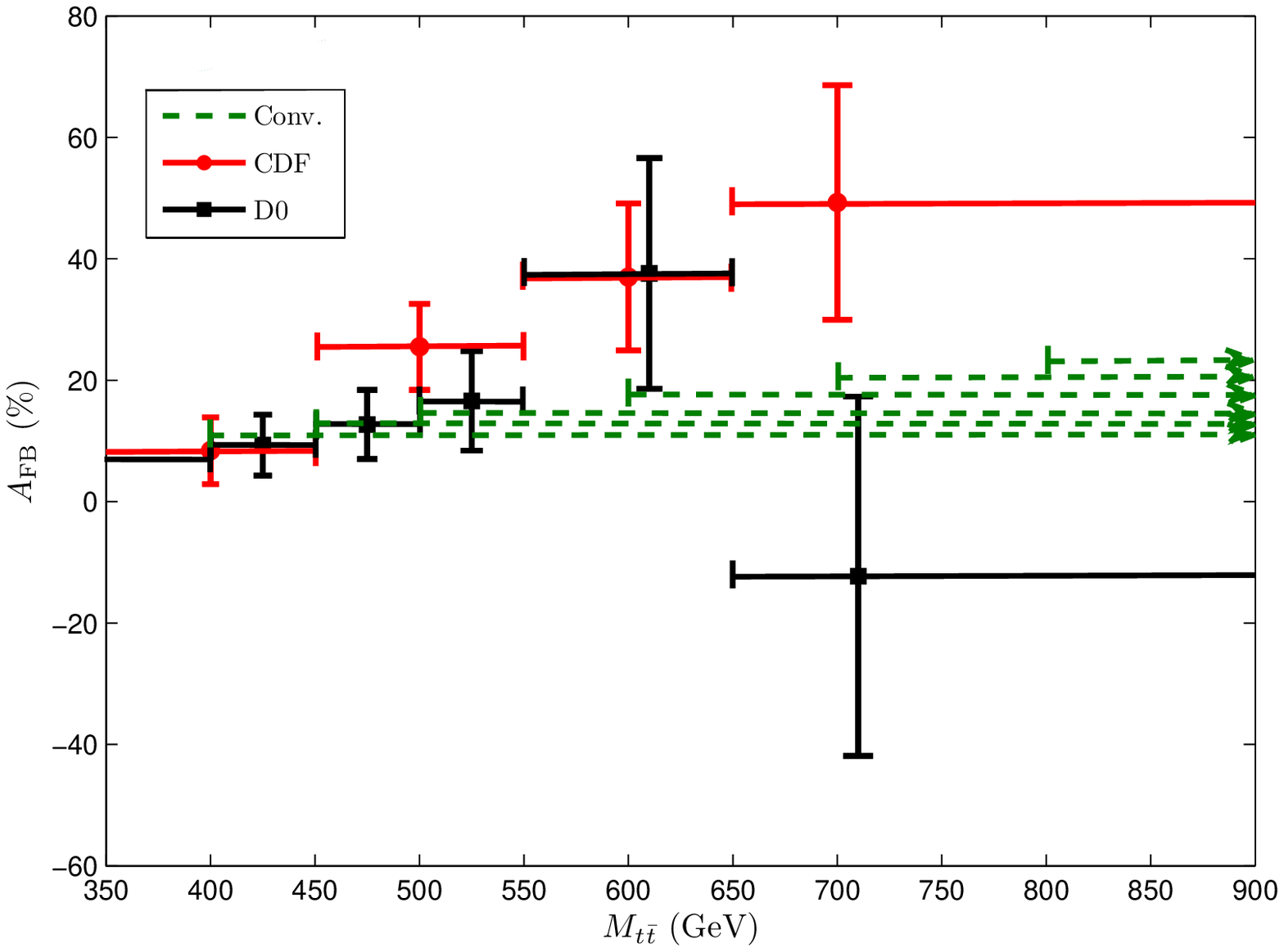}
\includegraphics[width=0.48\textwidth]{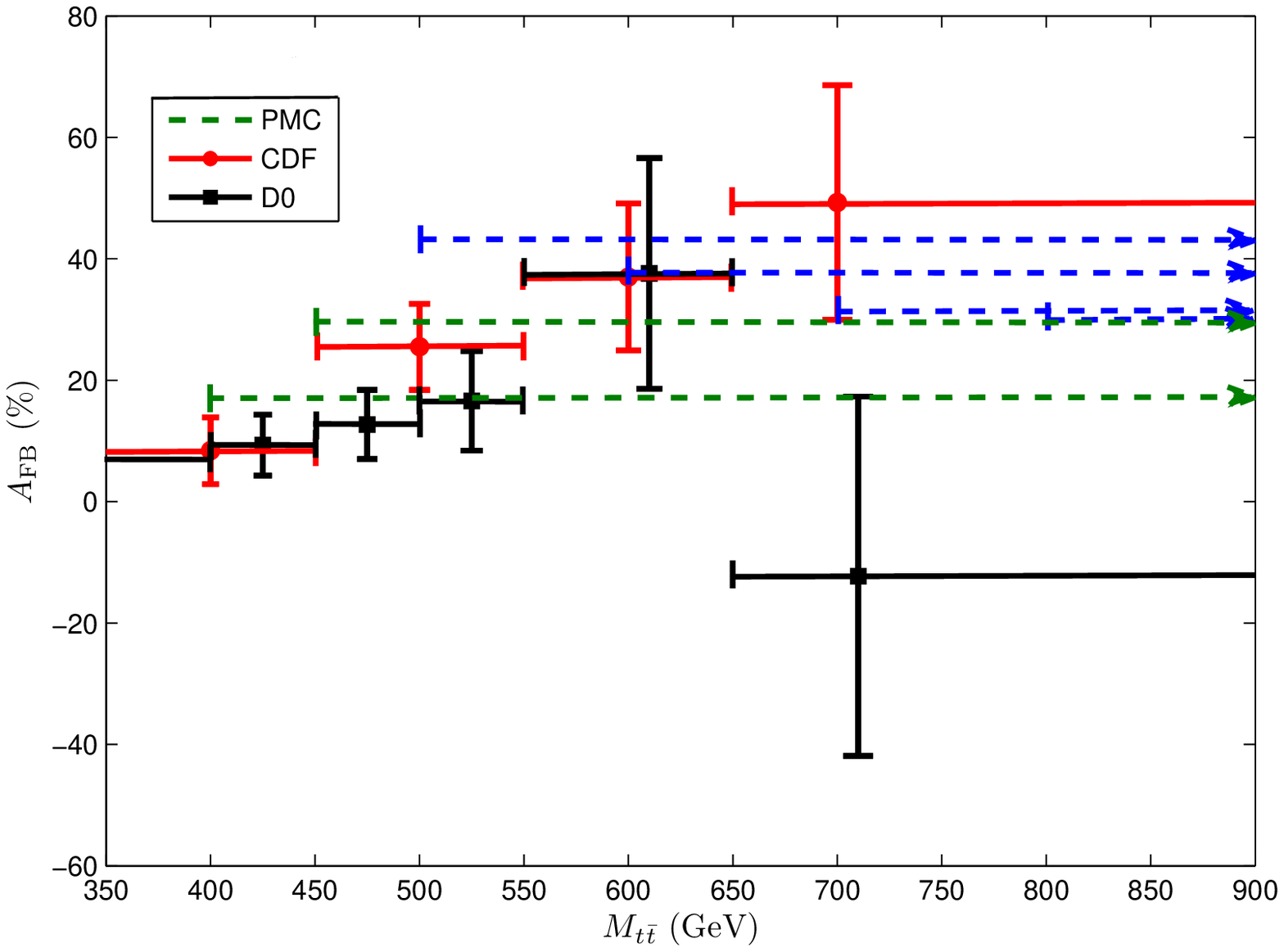}
\caption{A comparison of SM predictions of $A_{\rm FB}$ using conventional (Conv.) and PMC scale-settings with the CDF~\cite{Aaltonen:2012it} and D0~\cite{Abazov:2014cca} measurements~\cite{Wang:2015lna}. The initial scale $\mu_r=m_t$. } \label{fpmcafbetw}
\end{figure}
\end{center}
\end{widetext}

The most recent measurements reported by D0~\cite{Abazov:2014cca} indicate a non-monotonic, increasing-decreasing behavior for $A_{\rm FB}(M_{t\bar{t}}>M_{\rm cut})$ as the lower limit of the $t\bar t$ mass is increased. This behavior cannot be explained even by a NNLO QCD calculation using conventional scale-setting; one predicts monotonically increasing behavior~\cite{Czakon:2014xsa}. More explicitly, as shown in Table \ref{tab_mtcut}, if one assumes conventional scale-setting with the fixed scale $m_t$, then $A_{\rm FB}(M_{t\bar{t}}>M_{\rm cut})$ monotonically increases with increasing $M_{\rm cut}$. In contrast, if one employs the PMC, then $A_{\rm FB}(M_{t\bar{t}}>M_{\rm cut})$ first increases and then decreases as the lower limit of the pair mass $M_{\rm cut}$ is increased. These trends are more clearly shown in Fig.(\ref{fpmcafbetw}), in which the Standard Model predictions using conventional and PMC scale-settings are compared with the CDF~\cite{Aaltonen:2012it} and D0~\cite{Abazov:2014cca} measurements. The PMC predictions can be understood in terms of the behavior of the effective pQCD coupling $\bar\alpha_s(\overline{\mu}^{\rm PMC}_r)$; it is the weighted average of the running couplings entering the $(q\bar{q})$-channel, the subprocess underlying the asymmetry in pQCD. The effective coupling $\bar\alpha_s(\overline{\mu}^{\rm PMC}_r)$ depends in detail on kinematics, and the non-monotonic behavior of the effective coupling accounts for the ``increasing -decreasing" behavior of $A_{\rm FB}(M_{t\bar{t}}>M_{\rm cut})$~\cite{Wang:2015lna}. We have also recently shown~\cite{Wang:2014sua} that the PMC predictions are in agreement with the available ATLAS and CMS data~. Thus, the proper setting of the renormalization scale provides a consistent Standard Model explanation of the top-quark pair asymmetry measurements at both the Tevatron and LHC.

\subsection{The angular distributions of massive quarks and leptons close to threshold}

The PMC can also be applied to problems with multiple physical scales. The $t\bar{t}$-pair hadronic production already provides one of such examples at the Tevatron or LHC. In the case of the hard part at the two-loop level, we need to introduce two PMC scales. The Coulomb-type corrections in the threshold region are enhanced by factors of $\pi$, thus the terms which are proportional to $(\pi/v)$ or $(\pi/v)^2$ must be treated separately and an extra PMC scale has to be introduced, which is relatively soft for $v\to 0$~\cite{Brodsky:2012sz}. As another example, a BLM analysis of the angular distributions of massive quarks and leptons close to threshold has been done. It was shown that the subprocess $q \bar q \to Q\bar Q$ near the quark threshold involves not only the subprocess scale $\sqrt{\hat{s}} \sim 2M_Q$ but also the scale $v\sqrt{\hat{s}}$ which enters the Sudakov final-state corrections~\cite{Brodsky:1995ds}, where $v$ is the $Q \bar Q$ relative velocity. At this order, the BLM and PMC predictions are the same, so for this particular process, we can treat BLM and the PMC on an equal footing. More explicitly, we need to introduce two PMC scales for this process, one for the hard part and one for the Coulomb-type terms. This example also illustrates how to deal with multiple scale problems, which is relevant for processes that can be studied at a super $\tau$-charm factory or high intensity electro-proton accelerators with similar center-of-mass collision energy.

An important consequence of the heavy-quark kinematics is that the production angle of a heavy hadron follows the direction of the parent heavy quark. Close to threshold, in the limit $v \to 0$, the center-of-mass angular distribution for $e^+ e^- \to Q \bar{Q}$ is isotropic, a result of $S$-wave dominance. The small admixture of $P$-waves slightly above threshold provides a contribution $\propto v^2\cos\theta$. The angular distribution is measurable, and one can define the ``anisotropy" $A(v^2)$ of the process as
\begin{equation}
\frac{dN}{d\cos\theta}\propto 1+ A(v^2)\cos^2\theta.
\end{equation}
The anisotropy $A(v^2)$ can determined by the Dirac ($F_1$) and Pauli ($F_2$) form factors via the way~\cite{Brodsky:1995ds}
\begin{equation}
A=\frac{\tilde{A}}{1-\tilde{A}},
\end{equation}
with
\begin{equation}
\tilde{A}=\frac{v^2}{2} \frac{|F_1|^2(1-\beta^2)-|F_2|^2} {|F_1+F_2|^2(1-\beta^2)}.
\end{equation}
The two-loop-QED corrections to the form factors $F_{1,2}$ have been calculated in Ref.\cite{Hoang:1995ex}; thus one can set their renormalization scales by applying the PMC. For example, we have~\cite{Brodsky:1995ds}
\begin{eqnarray}
F_{1}+F_{2} &=& 1+\frac{\pi\alpha(v\sqrt{\hat{s}})} {4v}-2\frac{\alpha(\sqrt{\hat{s}} e^{3/8}/2)}{\pi} \\
&\simeq& \left(1-2\frac{\alpha(\sqrt{\hat{s}} e^{3/8}/2)}{\pi}\right) \left(1+\frac{\pi\alpha(\sqrt{\hat{s}} v)}{4v}\right). \label{FF12}
\end{eqnarray}
One finds two distinctly different correction factors. The first originates from hard transverse photon exchange, where the scale reflects the short distance process; the second arises from the instantaneous Coulomb potential. All of the $1/v$-terms can then be resummed using Sommerfeld's re-scattering formula. For example, from Eq.(\ref{FF12}), we can get
\begin{equation}
|F_1 + F_2|^2 \simeq \left( 1-4\frac{\alpha(\sqrt{\hat{s}} e^{3/8}/2)}{\pi}\right)\frac{x}{1-e^{-x}},
\end{equation}
where $x={\pi\alpha(\sqrt{\hat{s}} v)}/{v}$. One can take $\hat{s}\simeq 4m_Q^2$ in the threshold region. In this way one can predict $|F_1|^2$, $|F_2|^2$ and $|F_1 + F_2|^2$, and thus give accurate predictions for the anisotropy $A$. These formulae can be conveniently matched to the QCD case by using the effective charge of the potential $\alpha_V$~\cite{Brodsky:1995ds}. Because $A$ is sensitive to $\alpha_V(\sqrt{\hat{s}} v)$, the measurement of anisotropy can provide a check on other determinations of $\alpha_V$.

The anisotropy of $\tau$-pairs produced through the channel $e^+ e^-\to\tau^+ \tau^-$ can also be used in an analogous way to measure the Pauli form factor $F_2(\hat{s})$ of the $\tau$ lepton in the threshold domain $\hat{s} \geq 4m_\tau^2$. Thus a precise measurement of the anisotropy could provide a novel measurement of a fundamental parameter of the $\tau$ lepton and its time-like anomalous magnetic moment.

\section{Summary}

It is clearly important and fundamental to set the renormalization scale in a manner consistent with the principles of the renormalization group. The most critical criterion is that a prediction for a physical observable cannot depend on a theoretical convention such as the choice of renormalization scheme or the (initial) scale. This principle is satisfied by GM-L scale setting which is rigorously used for precision QED predictions. The QED scale is unambiguous, and the resulting high precision QED predictions are the same in any scheme at any finite order. The same properties are also satisfied for non-Abelian gauge theory when one uses PMC scale-setting. The PMC can be applied to any high-order process thus giving precision tests of theory at any experiment facility.

We have illustrated the main features of PMC predictions for hadronic $Z$ decays, the yields and the forward-backward asymmetry of the top-quark pair at the Tevatron, and the production of massive quarks and leptons close to threshold.

In contrast to predictions obtained using conventional scale setting, one finds for the PMC:

\begin{itemize}

\item All terms in the pQCD series involving the $\beta$-function are absorbed into the running coupling order-by-order. The value of the PMC scale at each order is fixed, which also determines the effective number of contributing flavors $n_f$ at each order, just as in QED. One finds negligible initial scale-independence of the PMC predictions for both the global observables and the differential observables at each order and each phase-space point. Small initial scale dependence can often also be achieved when using conventional scale setting at very high orders in pQCD; however, this is typically due to the cancelation among different orders or different phase-space points. Nevertheless, the scale dependences for differential observables at each order can remain very large. This fact underlies the inconsistency of conventional predictions of the top-quark pair asymmetry with measurements.

    When one applies the PMC, a large $t \bar t$ asymmetry is predicted, in agreement with data. This can be traced to the fact that a smaller effective PMC scale controls the NLO-terms of $q\bar{q}$-channel,  resulting in an enhanced NLO-contribution.  Furthermore, the NNLO terms are negligible due to the improvement of the pQCD convergence. The effective scale is determined by a NNLO $\beta_0$-term which is independent of the choice of initial scale; its behavior versus $M_{t\bar{t}}$ explains the ``increasing-decreasing" behavior measured by D0. The NNLO calculation of the top-pair asymmetry using conventional scale setting can also predict a reasonable large top-pair asymmetry. This is however due to a large contribution at NNLO. It has a large scale uncertainty, and it cannot explain the observed ``increasing-decreasing" behavior as a function of the $t \bar t$ invariant mass.

\item Only those $\beta$-terms which pertain to the running of $\alpha_s$  should be eliminated. One can confirm that the nonconformal $\beta$-terms are correctly identified and absorbed by the PMC procedure by checking that there is negligible dependence of the fixed-order theory prediction on the choice of the initial scale. Some dependence on the initial scale can persist using the PMC due to unknown higher-order terms; however, this dependence is highly suppressed even for low-order predictions.

\item The resulting coefficients of the pQCD series at any perturbative order using the PMC method are thus identical to that of the corresponding conformal theory with $\beta=0$, and the PMC predictions are thus scheme independent. Such scheme-independence is exact for all dimensional-like ${\cal R}_{\delta}$-schemes~\cite{Mojaza:2012mf, Brodsky:2013vpa}. One can also obtain scheme-independent predictions for effective charges using ``commensurate scale relations"~\cite{Brodsky:1994eh}. In principle, there can be some residual scheme dependence if one uses a non-dimensional-regularization scheme, due to unknown higher-order $\beta$-terms; in such a case, the PMC scale for the highest-known terms can only be determined by one-order-higher terms, which ay be unknown. Thus one cannot obtain a complete commensurate scale relation at this particular order. However, due to the elimination of the divergent renormalon terms, the value of the highest-known term itself is usually small, so such residual scheme dependence is typically highly suppressed.

\item PMC scale-setting also can be systematically applied to multi-scale problems. The typical momentum flow can be distinct; thus one should apply the PMC separately in each region. For example, in the case of the production massive quarks and leptons close to threshold, two PMC scales arise at NNLO~\cite{Brodsky:1995ds}; one is proportional to $\sqrt{\hat{s}}$ and the other one is proportional to $v\sqrt{\hat{s}}$.

    There are cases where additional momentum flows occur -- again contradicting conventional scale setting; however, the PMC eliminates such ambiguities. For example, there are two types of log-terms $\ln(\mu_r/M_{Z})$ and $\ln(\mu_r/M_{t})$ for the axial-singlet $r^A_S$ of the hadronic $Z$ decays. By applying the PMC, one finds the scale is $Q^{\rm AS} \simeq 100$ GeV~\cite{Wang:2014aqa}, indicating the typical momentum flow for $r^A_S$ is closer to $M_Z$ rather than $M_t$.
\end{itemize}

The Principle of Maximum Conformality is an important theoretical tool for making precise, reliable predictions for QCD. The PMC rigorously eliminates the usual ambiguities associated with the renormalization scale-setting procedure, giving predictions which are independent of the choice of the renormalization scheme. The scales of the QCD couplings and the effective number of quark flavors are systematically set, order-by-order, even for multiple-scale applications. The usual $n!$ divergent renormalon behavior of the perturbative expansion is also eliminated.

The application of the PMC can thus greatly improve the precision of tests of QCD at a Super Tau-Charm factory and the sensitivity of measurements at the LHC and other colliders to new physics beyond the Standard Model.  \\

\noindent{\bf Acknowledgments}: This review, submitted to {\it Frontiers of Physics}, is based on a contribution by S.J.B. at the Conference
{\it Workshop on Physics at a Future High Intensity Collider @ 2-7 GeV in China} Hefei, China January 14-16, 2015. This work was supported in part by the Natural Science Foundation of China under Grant No.11275280, the Department of Energy Contract No.DE-AC02-76SF00515, and by Fundamental Research Funds for the Central Universities under Grant No.CDJZR305513. SLAC-PUB-16357.

\end{document}